\documentclass[%
 reprint,
 amsmath,amssymb,
 aps,
]{revtex4-2}

\usepackage{graphicx}
\usepackage{dcolumn}
\usepackage{bm}
\usepackage{mathcomp}

\begin{document}

\preprint{APS/123-QED}

\title{Development of Nonlocal Kinetic-Energy Density Functional \\ for the Hybrid QM/MM Interaction }

\author{Hideaki Takahashi}
\email{ hideaki.takahashi.c4@tohoku.ac.jp}
\affiliation{%
Department of Chemistry, Graduate School of Science,\\
 Tohoku University, Sendai, Miyagi 980-8578, Japan
}%


\date{\today}

\begin{abstract}
Development of the electronic kinetic-energy density functional is a subject of major interest in theoretical physics and chemistry. In this work, the nonlocal kinetic-energy functional is developed in terms of the response function for the molecular system to realize the orbital free density-functional theory(OF-DFT) to be utilized in the hybrid QM/MM(quantum mechanical/molecular mechanical) method. The present approach shows a clear contrast to the previous functionals where the homogeneous electron gas serves as a reference to build the response function. As a benchmark test we apply the method to a QM water molecule in a dimer system and that embedded in a condensed environment to make comparisons with the results given by the QM/MM calculations employing the Kohn-Sham DFT. It was found that the energetics and the polarization density of the QM solute under the influence of the MM environment can be adequately reproduced with our approach. This work suggests the potential ability of the kinetic-energy functional based on the response functions for the molecular reference systems. 
  
\end{abstract}

\maketitle



Electronic density-functional theory (DFT) is an indispensable tool in modern physics and chemistry to study electronic structures of various materials\cite{rf:parr_yang_eng, Martin2004}. The extensive activities of DFT can be mainly attributed to the successful developments in the efficient and accurate density functionals for the exchange and correlation energies $E_{xc}$ of electrons\cite{Martin2004, Koch2001}. It has been fully established in many applications that the GGA (generalized gradient approximation) and the hybrid $E_{xc}$ functionals are capable of providing the electronic properties of the materials with accuracies comparable to those obtained by the highly sophisticated molecular orbitals theories within much less computational costs\cite{Koch2001}. 

It should be noted, however, that the rapid progress of the modern DFT is also dependent on the theoretical framework of the Kohn-Sham (KS) DFT\cite{rf:kohn1965pr} which utilizes the set of 1-electron wave functions $\{\varphi_i\}(i=1,2, \cdots, N)$ instead of the electron density $n$ to describe the kinetic energy of $N$ electrons in the system. The introduction of the wave functions is necessitated due to the difficulty in developing the accurate density functional $E_\text{kin}[n]$ for the kinetic energy\cite{rf:parr_yang_eng, Martin2004}. As a consequence, however, the computational cost inevitably increases due to the orthogonality conditions $\langle \varphi_i | \varphi_j \rangle = \delta_{ij} (i,j = 1,2, \cdots, N)$ imposed on the wave functions. Thus, the cost for KS-DFT calculation scales as $O(N^2)$ at least, which can be an obstacle in realizing the application of DFT to massively large systems.   

When an efficient kinetic-energy functional $E_\text{kin}[n]$ becomes available, it is possible to realize the computational costs to increase linearly with the system size since the wave functions are no longer needed. It is, thus, desirable to develop such a kinetic energy functional to extend the applicability of DFT to realistic systems\cite{Hung2009, rf:hung2010cpc}. Actually, substantial efforts have been devoted\cite{Wesolowski2013, Witt2018} to devise the functional $E_\text{kin}[n]$ whereby the orbital-free (OF) DFT calculation is made possible.         
 
Thomas and Fermi\cite{Thomas1927PCPS, Fermi1928zp} provided the first primitive approximation to $E_\text{kin}[n]$ by adopting the exact kinetic energy of the homogeneous electron gas (HEG) with density $n$ to the system of interest in the same way as the local density approximation (LDA). The correction to the TF approximation was provided by von Weizs\"{a}cker\cite{Weizsacker1935zp} by adding the gradient term $E_\text{vW}[n]$ to $E_\text{TF}[n]$. Unfortunately, it is known that higher order gradient corrections make quite minor contributions to improve the kinetic energy. It is considered in general that the local and the semilocal kinetic operators by themselves would not be able to reproduce the shell structures intrinsic to the electron densities in atoms. This is regarded as the critical deficiency in these functionals.            

On the basis of above discussion, Wang and Teter\cite{Wang1992prb} proposed an ansatz for a nonlocal kinetic energy functional in the form $E_{\text{WT}}^{\text{nloc}}[n]=C_{\text{WT}} \int d \bm{r} d \bm{r}^{\prime} n(\bm{r})^{\alpha} \omega_{0}\left(k_{F}\left|\bm{r}-\bm{r}^{\prime}\right|\right) n\left(\bm{r}^{\prime}\right)^{\alpha}$ where $\bm{r}$ and $\bm{r}^\prime$ are the coordinates of the electrons, and $\omega_{0}$ is constructed from the response function\cite{Lindhard1954} of HEG with density $n_0$ specified by the Fermi wavenumber $k_F = (3\pi^2n_0)^{1/3}$. The parameter $\alpha$ in $E_{\text{WT}}^{\text{nloc}}[n]$ was taken as $5/6$ in the original development. Importantly, it was demonstrated in Ref. \cite{Wang1992prb} that the incorporation of the nonlocal term $E_{\text{WT}}^{\text{nloc}}[n]$ into the energies $E_\text{TF}$ and $E_\text{vW}$ retrieves the shell structures in the production of atomic electron densities. Thus, the nonlocal term is regarded as an essential ingredient for the efficient kinetic energy functional.     

In this Letter, we develop a novel nonlocal kinetic energy functional for the purpose to evaluate the interactions in hybrid quantum mechanical / molecular mechanical(QM/MM) method\cite{rf:warshel1976jmb, rf:gao1992sci, rf:ruizlopez2003jmst, rf:takahashi2001jcc}, which enables ones to perform the orbital-free QM/MM simulations. A novel feature of the present approach is that a molecular system is employed instead of HEG as a reference for the response function which serves to construct the nonlocal kinetic energy functional. This shows a clear contrast to the previous developments\cite{Witt2018} of the functional $E_\text{kin}[n]$.    

In the following, we illustrate the outline of the formulation of our kinetic energy functional, followed by the method of its implementation on the real-space grid approach. We start the discussion with the definition for the total energy $E[n]$ of the system with an electron density $n$,
\begin{equation}
E[n]=E_{\text{kin}}[n]+E_{\text{H}}[n]+E_{xc}[n]+E_{\text{ext}}[n] \;\;\;  .
\label{eq:E_n}
\end{equation}
The second and fourth terms in the right hand side, that is, $E_{\text{H}}[n]$ and $E_{\text{ext}}[n]$, represent the classical Coulomb energy among the electrons of the density $n$ and the interaction between the external potential $\upsilon_\text{ext}(\bm{r})$ and the electrons, respectively. The first derivative of $E[n]$ with respect to $n(\bm{r})$ at its equilibrium density $n_0$ gives the chemical potential $\mu$ of the electrons,  
\begin{equation}
\left. \frac{\delta E[n]}{\delta n(\bm{r})}\right|_{n=n_0} = \upsilon_\text{kin}[n_0](\bm{r}) + \upsilon_\text{eff}[n_0](\bm{r}) = \mu
\label{eq:dEndn}
\end{equation}
where we define $\upsilon_\text{kin}[n](\bm{r}) = \frac{\delta E_\text{kin}[n]}{\delta n(\bm{r})}$ and $\upsilon_\text{eff}[n](\bm{r}) = \frac{\delta}{\delta n(\bm{r})}(E_{\text{H}}[n]+E_{xc}[n]+E_{\text{ext}}[n])$. Then, the second derivative of $E[n]$ around the stationary density becomes zero, 
\begin{align}
& \left. \frac{\delta^{2} E[n]} {\delta n(\bm{r}) \delta n\left(\bm{r}^{\prime}\right)} \right|_{n=n_0}   \notag  \\
& = \left. \frac{\delta \upsilon_{\text{kin}}[n](\bm{r})}{\delta n\left(\bm{r}^{\prime}\right)} \right|_{n=n_0} +  \left. \frac{\delta \upsilon_{\text{eff}}[n](\bm{r})}{\delta n\left(\bm{r}^{\prime}\right)}  \right|_{n=n_0} =0
\end{align}
Then, it is possible to relate the derivative of the kinetic potential $\upsilon_\text{kin}[n](\bm{r})$ to the response function $\chi (\bm{r}, \bm{r}^{\prime})$ of the system, 
\begin{align}
 \left. \frac{\delta \upsilon_{\text {kin}}[n](\bm{r})}{\delta n\left(\bm{r}^{\prime}\right)}  \right|_{n=n_0}  =  \left. -\frac{\delta \upsilon_{\text {eff}}[n](\bm{r})}{\delta n\left(\bm{r}^{\prime}\right)} \right|_{n=n_0}  
  =-\chi_0^{-1}\left(\bm{r}, \bm{r}^{\prime}\right)
\end{align}

Now we introduce an approximation to $E_\text{kin}[n]$ for a certain density $n$ through the second-order Taylor expansion around the density $n_0$
\begin{align} 
E_\text{kin}[n] = & E_\text{kin}[n_0] + \int d\bm{r} \upsilon_\text{kin}[n_0](\bm{r})\delta n(\bm{r})   \notag  \\
& -\frac{1}{2} \int d\bm{r} d\bm{r}^\prime \delta n(\bm{r}) \chi_0^{-1}\left(\bm{r}, \bm{r}^{\prime}\right) \delta n(\bm{r}^\prime)
\label{eq:Ekin}  
\end{align}
where $\delta n$ is defined as $\delta n = n - n_0$. Similarly, the kinetic potential $\upsilon_\text{kin}[n](\bm{r})$ is expressed as
\begin{equation}
\upsilon_\text{kin}[n](\bm{r}) = \upsilon_\text{kin}[n_0](\bm{r}) - \int d\bm{r}^\prime \chi_0^{-1}\left(\bm{r}, \bm{r}^{\prime}\right) \delta n(\bm{r}^\prime)
\end{equation}

Considering that the present method is to be employed e.g. in QM/MM simulations, $n_0$ should be naturally taken as the electron density of the QM molecule at its isolation, while $n$ is that under the influence of the Coulomb potential $\upsilon_\text{MM}(\bm{r})$ created by the MM environment. Thus, the reference density $n_0(\bm{r})$ is obtained from the solution of the following KS equation for the isolated QM system,
\begin{equation}   
\left( -\frac{1}{2} \nabla^2 + \upsilon_\text{eff}[n_0](\bm{r}) \right)  \varphi_i^0(\bm{r})  = \epsilon_i^0 \varphi_i^0(\bm{r})  
\label{eq:KS_n0}
\end{equation}
where $n_0(\bm{r})$ is given by $n_0(\bm{r}) = \sum_i^N |\varphi_i^0(\bm{r})|^2$. The response function $\chi_0(\bm{r}, \bm{r}^\prime)$ can also be expressed in terms of the occupied and virtual orbitals $\{\varphi_i^0\}$ in Eq. (\ref{eq:KS_n0}) through the second-order perturbation theory, 
\begin{align} 
\chi_{0}\left(\bm{r}, \bm{r}^{\prime}\right)=\sum_{i}^{\text {occ}} \sum_{a}^{\text {vir}} \frac{1}{\varepsilon_{a}^{0}-\varepsilon_{i}^{0}} & \varphi_{i}^{0 *}(\bm{r}) \varphi_{a}^{0}(\bm{r})    \notag   \\
&\times \varphi_{a}^{0 *}\left(\bm{r}^{\prime}\right) \varphi_{i}^{0}\left(\bm{r}^{\prime}\right)
\label{eq:PT2}
\end{align}
Once the solution of the Eq. (\ref{eq:KS_n0}) is obtained for the isolated QM system, it is possible to evaluate the kinetic energy $E_\text{kin}[n]$ of the system perturbed e.g. by the MM potential $\upsilon_\text{MM}(\bm{r})$ through Eq. (\ref{eq:Ekin}) with accuracy up to the second order of $\delta n$. The kinetic energy functional $E_\text{kin}$ was, thus, formulated on the basis of the molecular system as a reference system.   

It should be noted, however, that the numerical implementation of the above formulation is not straightforward in general. Actually, it seems computationally infeasible to make an inversion of the six-dimensional response function $\chi_{0}\left(\bm{r}, \bm{r}^{\prime}\right)$. Below, we provide a practical solution to perform the numerical inversion of the matrix. First, we introduce a certain orthonormalized set of  basis functions $\{\psi_i \}$ to represent the response operator $\hat \chi_0$, i.e. $(\chi_0)_{ij} = \langle \psi_i | \hat \chi_0 | \psi_j \rangle$. Assuming that a spectral decomposition is provided for $\hat \chi_0$ as $\hat \chi_{0}=\sum_{p}^{N_s} \left. \mid \gamma_{p}\right\rangle g_{p}\left\langle\gamma_{p} \mid \right. $, then, $(\chi_0)_{ij}$ is
represented as 
\begin{equation} 
(\chi_0)_{ij} = \sum_{p}^{N_s} \langle \psi_i \mid \gamma_{p} \rangle g_{p} \langle\gamma_{p} \mid  \psi_j \rangle
\label{eq:spec_decmp}
\end{equation}
where it is readily recognized that $\langle \psi_i | \gamma_{p} \rangle$ is the matrix element of the unitary transformation $U$ that diagonalizes the matrix $(\chi_0)_{ij}$. Our present purpose is to construct the eigenvectors $|\gamma_p \rangle$ with the given basis functions $|\psi_i \rangle$. Since $(\chi_0)_{ij}$ can be computed using Eq. (\ref{eq:PT2}), the unitary matrix $U$ can also be provided. $\gamma_p(\bm{r})$ is, then, transformed to 
$\langle \bm{r} | \gamma_p \rangle = \sum_i^{N_s} \langle \bm{r} | \psi_{i} \rangle  U_{ip} $. With $\gamma_p(\bm{r})$ and the corresponding eigenvalues $g_i$ thus obtained, the inversion of the matrix $\hat \chi_0$ is given by $\chi_0^{-1}(\bm{r}, \bm{r}^\prime) = \sum_{p}^{N_s} \langle \bm{r} | \gamma_{p} \rangle g_{p}^{-1} \langle\gamma_{p} | \bm{r}^\prime \rangle $. In our implementation, we employ the eigenvectors $\{\varphi_i^0\}$ of Eq. (\ref{eq:KS_n0}) as the basis functions. We must take care, however, in inverting the matrix $\hat \chi_0$, since the inverse of a small eigenvalue of $g_p$ will lead to a numerical instability.  This will be later discussed when we apply our method to compute the hydrogen bond energy in a water dimer. Anyway, the numerical method to evaluate the nonlocal term in Eq. (\ref{eq:Ekin}) is thus obtained. 

In the evaluation of the second term in the right hand side of Eq. (\ref{eq:Ekin}), we take advantage of Eq. (\ref{eq:dEndn}). Since it is naturally assumed that some adequate $E_{xc}[n]$ functional is available, $\upsilon_{\text{eff}}[n_0](\bm{r})$ can be readily evaluated. That is, $\upsilon_\text{kin}[n_0](\bm{r})$ is to be evaluated as $\upsilon_{\text{kin}}[n_0](\bm{r})= \mu -\upsilon_{\text{eff}}[n_0](\bm{r})$. The constant shift of the potential by the chemical potential $\mu$ of the electrons can be taken arbitrarily since $\int d\bm{r} \delta n(\bm{r}) = 0$ holds exactly. Thus, we take the negative sign of $\upsilon_{\text{eff}}[n_0](\bm{r})$ as the kinetic potential $\upsilon_{\text{kin}}[n_0](\bm{r})$. In the present implementation, however, we employ the real-space grid approach\cite{rf:chelikowsky1994prb,rf:chelikowsky1994prl,rf:takahashi2000cl,rf:takahashi2001jcc,takahashi2001jpca} to represent the electron density $n_0$ as well as the wave functions, which necessitates the use of nonlocal pseudopotentials\cite{Martin2004} for the electron-nuclei potential $\upsilon_\text{ext}(\bm{r})$. As a consequence, the kinetic potential $\upsilon_\text{kin}[n_0]$ becomes also dependent on the wave functions. This issue also applies to the calculation with a plane wave basis set. To bypass such an unfavorable situation, we utilize the reverse Kohn-Sham method\cite{rf:leeuwen1994pra} whereby the local potential $\upsilon_\text{eff}[n_0](\bm{r})$ corresponding to $n_0$ can be generated.    
      
\begin{figure}[h]
\centering
\scalebox{0.45}[0.45] {\includegraphics[trim=140 120 120 130,clip]{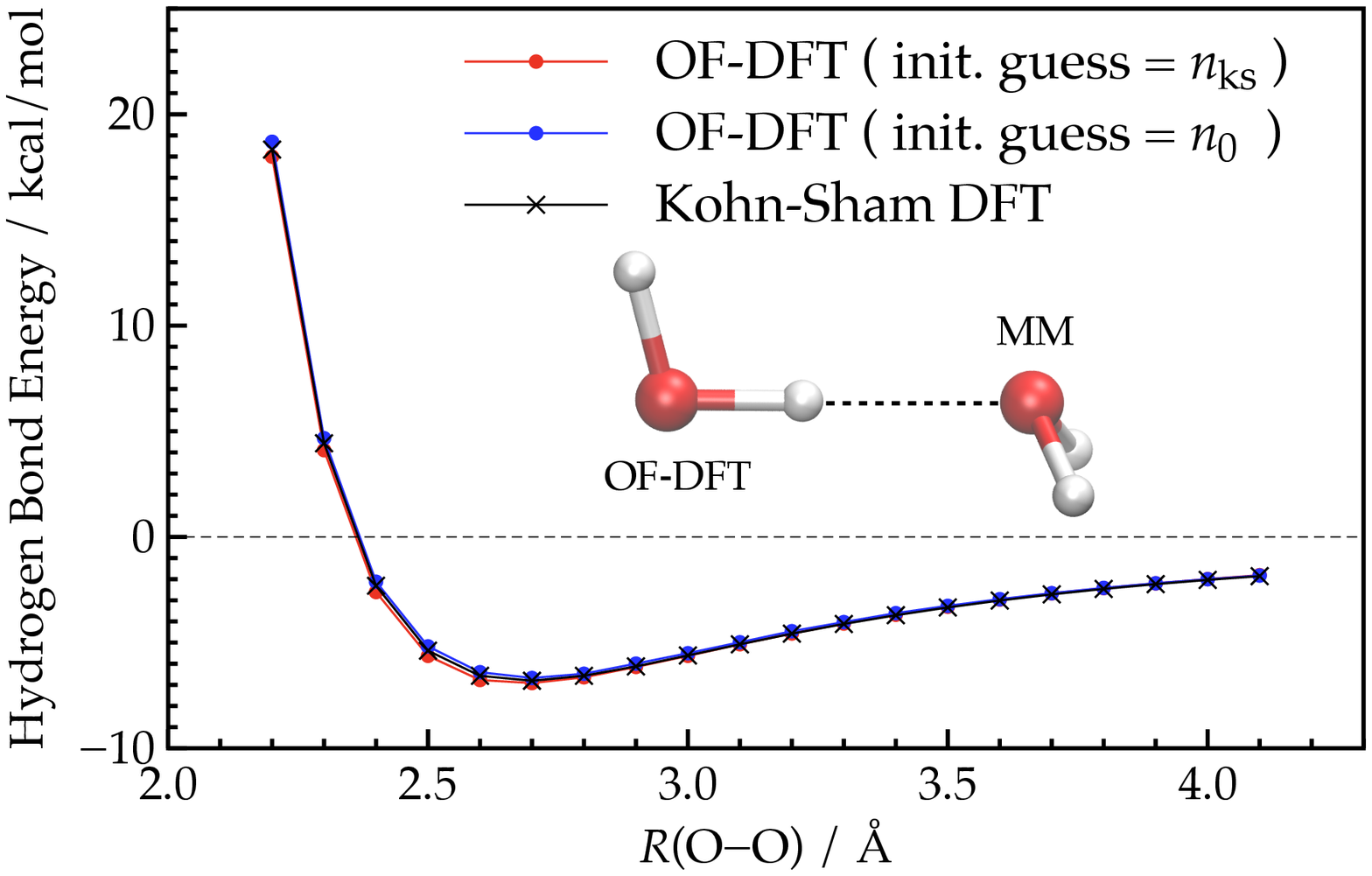}}            
\caption{\label{HB}Hydrogen bond energy curves of a water dimer as functions of the distance $R$(O$-$O) of the oxygen atoms. The proton donor molecule is described with the orbital-free DFT or Kohn-Sham DFT method, while the accepter is expressed by the SPC/E model. The molecular plane of the proton accepter is slanted by $60^\circ$ with respect to the O-O axis. The structure of each water molecule is being fixed for the variation of the O-O distance( bond length of $R$(O-H) $= 0.963$ \AA\; and the angle $\angle$HOH $= 104.6^\circ$). }
\end{figure}

We now provide the outline of the numerical matters to perform the OF-QM/MM calculation. The wave functions to construct the reference density $n_0$ are represented on the real-space grids\cite{rf:chelikowsky1994prb,rf:chelikowsky1994prl,rf:takahashi2000cl,rf:takahashi2001jcc,takahashi2001jpca} augmented with the double-grid technique\cite{rf:ono1999prl}. The width $h$ of the coarse grid is $0.248$ a.u. and that of the dense grid is set at $h/5$. The number of the coarse grids along each axis of the cubic QM cell is $80$, which leads to the box size $l = 19.8$ a.u. The exchange-correlation energy $E_{xc}$ is evaluated by the BLYP functional\cite{rf:becke1988pra,rf:lee1988prb}.  The nonlocal pseudopotentials with the form proposed by Kleinman and Bylander\cite{rf:kleinman1982prl}  are utilized to ensure the smooth behavior of the density around the atomic cores. In computing the hydrogen bond energy of the water dimer, the proton donor is regarded as the QM molecule and described with the present OF-DFT method, while the accepter molecule is represented with the SPC/E model\cite{rf:berendsen1987jpc} of water. The internal geometry of each water molecule is that optimized by the KS-DFT method with the B3LYP functional \cite{rf:becke1988pra,rf:becke1993Ajcp, rf:lee1988prb} and aug-cc-pVTZ basis set. The hydrogen-bonded complex has the structure of $C_{s}$ symmetry,  where a hydrogen atom of the donor molecule is placed on the line connecting the two oxygen atoms. The illustration of the complex is superposed in Fig. \ref{HB} for reference.     

When the density is converged at $n$ through a self-consistent field (SCF) calculation, it is guaranteed that the potential $\upsilon_\text{OF-DFT}[n]$ of Eq. (\ref{eq:dEndn}) becomes zero 
\begin{equation} 
\upsilon_\text{OF-DFT}[n](\bm{r}) = \upsilon_\text{kin}[n](\bm{r}) + \upsilon_\text{eff}[n](\bm{r}) = 0
\label{eq:SCF}
\end{equation}
provided that the standard of the potential is shifted by $-\mu$. The explicit form of the potential $\upsilon_\text{OF-DFT}[n]$ is given by 
\begin{align} 
\upsilon_\text{OF-DFT}&[n](\bm{r})   \notag  \\
 = \int & d\bm{r}^\prime \chi_0^{-1}(\bm{r},\bm{r}^\prime)\delta n(\bm{r}^\prime)
 + \upsilon_\text{H}[n](\bm{r}) - \upsilon_\text{H}[n_0](\bm{r})  \notag  \\
& + \upsilon_{xc}[n](\bm{r}) - \upsilon_{xc}[n_0](\bm{r}) + \upsilon_\text{MM}(\bm{r})  
\label{eq:v_ofdft}
\end{align}
Note that the right hand side in Eq. (\ref{eq:v_ofdft}) includes the external Coulomb potential $\upsilon_\text{MM}$ due to the point charges $\{q_k\}$ placed on the sites $\{\bm{x}_k\}$ in the MM molecule. $\upsilon_\text{MM}$ is merely given by $\upsilon_\text{MM}(\bm{r}) = \sum_k^\text{MM} q_k/| \bm{r}-\bm{x}_k | $. In our implementation the singularities at $\bm{x}_k$ are avoided by the method provided in Ref. \cite{rf:takahashi2001jcc}. The electron density $n$ is deformed from the reference distribution $n_0$ due to the potential $\upsilon_\text{MM}(\bm{r})$. The update of the density in the SCF procedure to realize Eq. (\ref{eq:SCF}) is expedited by means of the algorithm based on the Goldstein-Armijo principle\cite{Gill1981}. 

\begin{figure}[h]
\centering
\scalebox{0.50}[0.50] {\includegraphics[trim=170 190 120 130,clip]{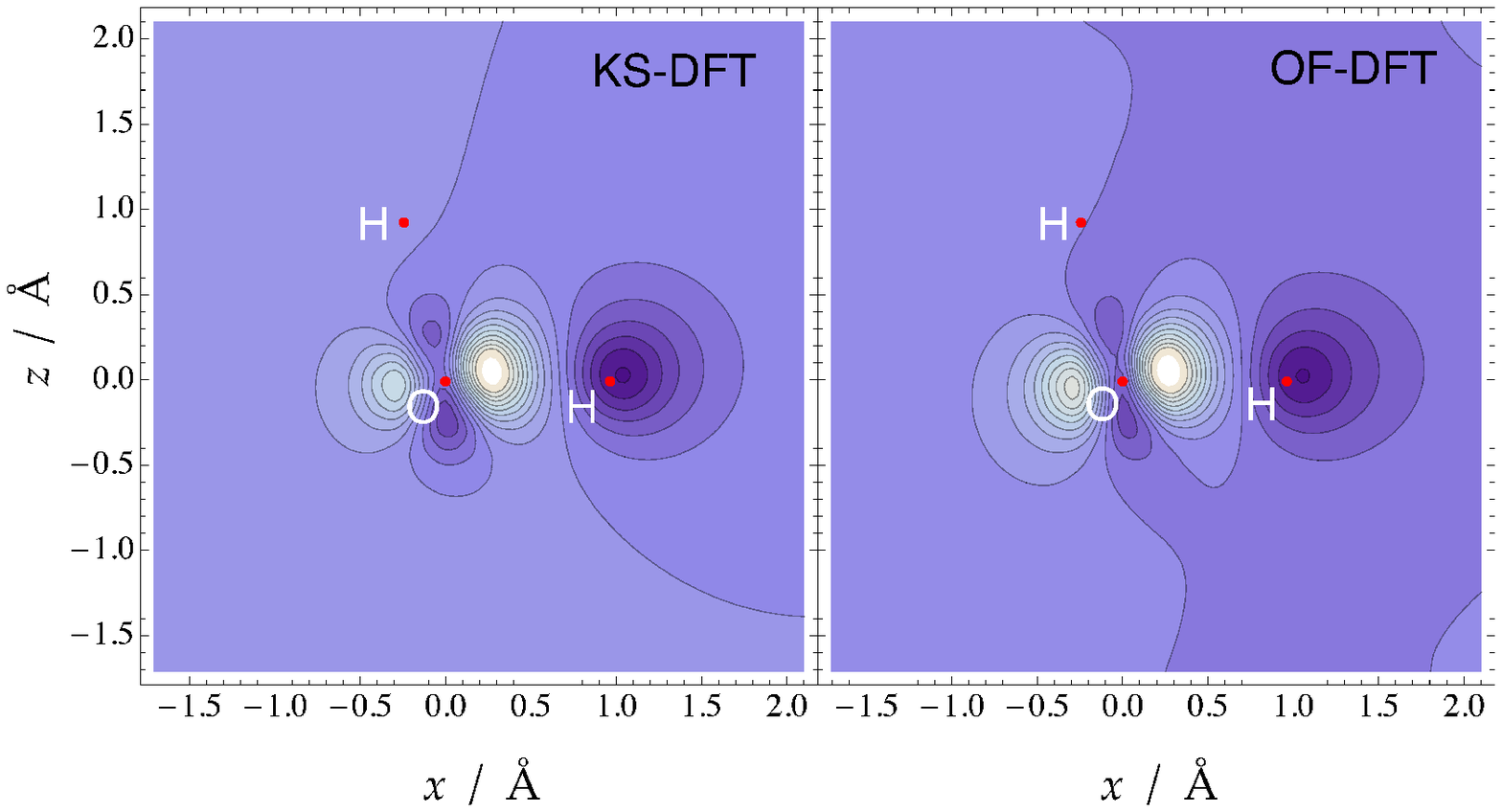}}            
\caption{\label{drho}The polarization densities $\delta n(\bm{r}) = n(\bm{r})-n_0(\bm{r})$ on the molecular plane of the proton donor in Fig. \ref{HB} with $R$(O$-$O) $= 2.6$ \AA. Figure on the left is $\delta n$ by KS-DFT and right is that by OF-DFT. The atoms in the water molecule are depicted with the red dots in the $x-z$ plane.  The variation of $\delta n(\bm{r})$ within the range of $\pm 0.02$ a.u.$^{-3}$ is plotted with 20 contours (lighter and darker colors denote the increase and decrease in the density, respectively. ) }
\end{figure}

Figure \ref{HB} shows the hydrogen bond energy curves of the water dimer obtained by present OF-DFT method,  where we make comparisons with that computed by KS-DFT. We also examine the dependence of the SCF convergence on the initial guess for the density. To this end we test two initial guesses, that is, the electron density $n_0$ of the QM solute at isolation, and the density $n_\text{KS}$ optimized by KS-DFT calculation. It is clearly shown in the figure that the curves by the present OF-DFT approach nicely agree with that produced by KS-DFT although the initial guess of $n_0$ gives the slightly larger energy than that of $n_\text{KS}$ especially in the small $R\text{(O-O)}$ region. It is, thus, demonstrated that the present OF-DFT method is able to compute the QM/MM interaction with the accuracy comparable to the KS-DFT. Throughout the present application, the eigenvectors $\gamma_p(\bm{r})$ in Eq. (\ref{eq:spec_decmp}) with their eigenvalues $g_p$ larger than $\varepsilon_\text{cut} = 10^{-5}$ a.u.$^{-1}$ were included in the construction of the response function. It was found, however, that the dimer energy at $R\text{(O-O)}= 2.6$ \AA\; is decreased by $0.24$ kcal/mol when $\varepsilon_\text{cut}$ is changed to $10^{-4}$ a.u.$^{-1}$. It is, thus, revealed that the energetics is not seriously dependent on the choice of the threshold $\varepsilon_\text{cut}$. However, a certain care must be taken for $\varepsilon_\text{cut}$ to achieve the desired accuracy.   We also examine the polarization density $\delta n(\bm{r})$ induced in the QM molecule in the water dimer. In Fig. \ref{drho} $\delta n$ computed by OF-DFT is compared with that given by KS-DFT. Although the details of the behavior of $\delta n$ in OF-DFT somewhat differs from that KS-DFT, it is found that the overall polarization can be soundly reproduced by the present OF-DFT. It is seen in the figure that the density is decreased on H facing the accepter molecule, while the population on the oxygen is increased. Obviously, such a polarization is caused by the excitation of the electrons to the antibonding $\sigma^*$ orbital composed of $s$ and $p_x$ atomic orbitals on the hydrogen and oxygen atoms, respectively. It is thus demonstrated that the present OF-DFT approach is capable of describing reasonably the spatial variation of the polarization density.        

\begin{figure}[h]
\centering
\scalebox{0.49}[0.49] {\includegraphics[trim=180 190 120 160,clip]{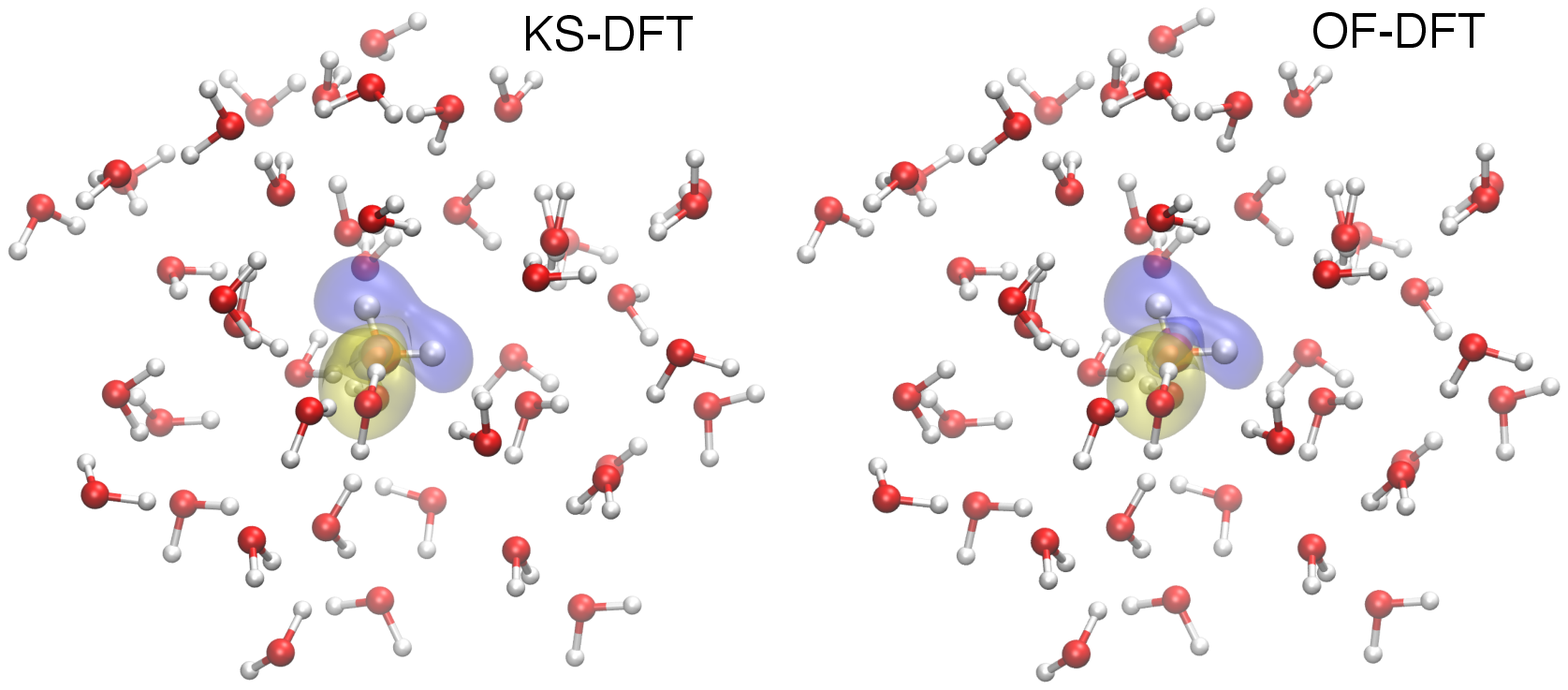}}            
\caption{\label{cluster}The ball-and-stick illustration of the water cluster employed in the present QM/MM calculation (left: KS-DFT, right: OF-DFT), where the oxygen atoms of water are depicted with the red spheres. The polarization densities $\delta n(\bm{r}) = n(\bm{r})-n_0(\bm{r})$ of the QM water molecule are also represented with the blue and yellow isosurfaces(blue: $\delta n = -0.001$ a.u.$^{-3}$, yellow: $\delta n = 0.001$ a.u.$^{-3}$). }
\end{figure}

We are also interested in the energetics of a QM molecule placed in a condensed environment. To this end, we consider a QM water embedded in a water cluster described by the SPC/E model. The structure of the cluster is that picked out arbitrarily from a trajectory in a simulation with 500 water molecules under the thermodynamic conditions $T = 298.15$ K and $\rho = 1.0$ g/cm$^3$.  In the calculations we consider the water molecules of which $R$(O-O) distances from the solute are smaller than $7.0$ \AA. As a result 47 MM water molecules are included in the cluster. The illustration of the cluster is shown in Fig. \ref{cluster}, where the isosurfaces of the polarization density provided by the OF-DFT method are compared with those obtained by the KS-DFT. The polarization provided in the OF-DFT calculation shows the good agreement with that given by the KS-DFT as demonstrated similarly in the dimer system. The QM/MM interaction energy($= E_\text{QM}+E_\text{QM/MM} - E_0$) is evaluated as $-28.4$ kcal/mol by the KS-DFT, which shows excellent agreement with the value $-28.8$ kcal/mol obtained by OF-DFT which employs the density $n_\text{KS}$ as the initial guess. We note, however, that when the density $n_0$ is used as the initial guess, the interaction energy is reduced to $-27.3$ kcal/mol, which shows the similar trend found in the calculation for the water dimer. Anyway, it is shown that the energetics in the condensed environment can also be evaluated with the accuracy comparable to the KS-DFT. 

Lastly, we make a comparison in efficiency between the present OF-QM/MM approach and the QM/MM based on the second-order perturbation theory (PT2)\cite{suzuoka2014jcp}. Since the PT2 approach also necessitates the orbitals and the corresponding eigenvalues to construct the energy and the polarization density, the computational cost for PT2 is almost equivalent to that for the present approach. Thus, we are interested in the accuracy of the OF-DFT method as compared to the PT2 calculation. In our previous work\cite{suzuoka2014jcp}, we performed a series of QM/MM-PT2 simulations combined with the free energy calculations for various molecules, where it was revealed that the PT2 approach fails to reproduce the solvation free energy of Acetonitrile (CH$_3$CN) into water solvent as compared to the QM/MM based on KS-DFT. It was concluded in the work that the failure can be attributed to the large interaction between the solute and solvent molecules due to the large permanent dipole moment of the molecule. Actually, by our KS-DFT calculation, the dipole is computed as 4.01 Debye. Thus, the evaluation of the interaction energy between Acetonitrile and water solvent is considered to be one of the good benchmark test for OF-DFT. The configuration of the water solvent is taken from the trajectory produced in the previous work. We refer the readers to Ref. \cite{suzuoka2014jcp} for the details of the QM/MM simulation. The setup for the present real-space grid calculation is also common to that for Ref. \cite{suzuoka2014jcp}. Anyway, 498 SPC/E water molecules are considered in the computation of the QM/MM interaction energies. As was done in the water dimer calculations, the local potential $\upsilon_\text{loc}[n_0](\bm{r})$ is also provided for Acetonitrile by the reverse Kohn-Sham method, which is followed by the construction of the response function $\chi_0$. The threshold for the eigenvalue $g_p$ in Eq. (\ref{eq:spec_decmp}) is set at $10^{-5}$ a.u.$^{-1}$ for the construction of $\chi_0$. The PT2 energy is evaluated using the orbitals and their eigenvalues constructed on the potential $\upsilon_\text{loc}[n_0](\bm{r})$, which differs from the work in Ref. \cite{suzuoka2014jcp}. 27 virtual orbitals for each spin are employed in the PT2 calculation. The interaction energy between Acetonitrile and the water solvent with a certain configuration is evaluated as $-43.1$ kcal/mol by KS-DFT, while it is obtained as $-46.3$ kcal/mol by the OF-DFT which employs $n_\text{KS}$ as the initial guess. Thus, the energy difference is obtained as $-3.2$ kcal/mol and it is slightly larger than the error for the QM water cluster. We note, however, that the difference in the interaction energy between the PT2 and the KS-DFT calculations becomes $-12.6$ kcal/mol. It is, thus, demonstrated that the OF-DFT offers an improvement in the energetics of the Acetonitrile embedded in water. This is probably because the present OF-DFT method has an advantage that the density can be optimized through SCF in contrast to the PT2 approach.   

In this work we developed a new method for OF-DFT that can be incorporated in the QM/MM approach. The notable feature of the method is that the nonlocal term in the kinetic energy functional is described with the inverse of the response function of the molecular system, which shows a clear contrast to the previous functional based on the response function for the homogeneous electron gas. It was demonstrated that the OF-DFT method is able to produce the QM/MM hydrogen bond energy curve of the water dimer in excellent agreement with that given by the QM/MM based on KS-DFT. The polarization density obtained by OF-DFT of the QM molecule in the water dimer also showed reasonable agreement with that given by KS-DFT. The energetics of an Acetonitrile as well as a water molecule in the condensed environment were also examined and the accuracy of the OF-DFT in the QM/MM simulation was confirmed. In the forthcoming issue, we will apply the present OF-QM/MM approach to the computations of the statistical properties of QM solutes in solutions such as solvation free energy.  
   
\begin{acknowledgments}
This paper was supported by the Grant-in-Aid for Scientific Research(C) (No. 17K05138, No. 22K12055) from the Japan Society for the Promotion of Science (JSPS); the Grant-in-Aid for Scientific Research on Innovative Areas (No. 23118701) from the Ministry of Education, Culture, Sports, Science, and Technology (MEXT); the Grant-in-Aid for Challenging Exploratory Research (No. 25620004) from the Japan Society for the Promotion of Science (JSPS). This research also used computational resources of the HPCI system provided by Kyoto, Nagoya, and Osaka university through the HPCI System Research Project (Project IDs: hp170046, hp180030, hp180032, hp190011, and hp200016).
\end{acknowledgments}  


\providecommand{\noopsort}[1]{}\providecommand{\singleletter}[1]{#1}%

\end{document}